\documentstyle[prb,preprint,aps]{revtex}    

\begin{document}
\draft
\newcommand{\beq}{\begin{equation}}
\newcommand{\eeq}{\end{equation}}
\newcommand{\beqa}{\begin{eqnarray}}
\newcommand{\eeqa}{\end{eqnarray}}

\title{Crossover from the Josephson effect \\ to bulk
superconducting flow}

\author{Fernando Sols and Jaime Ferrer}
\bigskip
\address{
Departamento de F\'{\i}sica de la Materia Condensada, C-XII\\
Universidad Aut\'onoma de Madrid, E-28049 Madrid, Spain}

\maketitle
\begin{abstract}
The crossover between ideal Josephson behavior and uniform superconducting
flow is studied by solving exactly
the Ginzburg-Landau equation for
a one-dimensional superconductor
in the presence of an effective delta function
potential of arbitrary strength.
As the effective scattering is turned off, the
pairs of Josephson solutions with equal current
evolve into a uniform and a
solitonic solution with nonzero phase offset.
It is also argued that a microscopic description of the crossover
must satisfy the self-consistency condition,
which is shown to guarantee current
conservation.
The adiabatic response to an external bias is briefly described.
The ac Josephson effect is shown to break down when the
external voltage is applied at points which are
sufficiently far from the junction.
\end{abstract}
\vspace{.5cm}


\vspace{.5cm}
\narrowtext

\section{Introduction}

The Josephson effect between two weakly coupled superconductors and the
steady flow of supercurrent in a perfect lead constitute the two main
paradigms of superconducting transport.
Both regimes can be viewed as the limits
of a general scenario in which Cooper pairs
flow coherently  in the presence of a scattering
obstacle of arbitrary strength.
The Josephson effect corresponds to the limit in
which a strongly reflecting obstacle
(typically, a tunneling barrier \cite{josephson} or a
point contact \cite{likharev})
reduces drastically the effective coupling between two
bulk superconductors while still preserving global coherence.
In the absence of an external bias, the current is given by the
Josephson relation
$I=I_C\sin(\Delta \varphi)$,
where $\Delta \varphi$ is the phase difference
between the two superconductors. The opposite limit is that of
supercurrent flow in
a perfect lead without appreciable scattering. In the appropiate gauge,
this regime is characterized
by a superconducting gap of uniform amplitude and
a linearly varying phase
whose gradient is proportional to the current. Specifically,
in the Ginzburg-Landau
limit, the current density can be written
$j=(e\hbar/m)|\psi|^2 \nabla
\varphi$, where $\psi=|\psi|e^{i\varphi}$ is the superconducting
order parameter.

An adequate measure of
the scattering strength is the average transmission
probability $T_0$ for a Fermi electron
passing through the barrier or contact in the normal phase,
\begin{equation}
T_0\equiv (h/e^2R_N)(2\pi/Ak_F^2)
\end{equation}
where $R_N$ is the device normal resistance, $A$ is the
cross section area of the
semiinfinite leads, and $k_F$ is the Fermi wave vector.
``Weak" and
``strong" superconductivities
are then characterized by $T_0 \ll 1$ and $T_0
\simeq 1$, respectively.
For a structure in which superconductivity is not weakened by
one-electron reflection, such as a {\it S-N-S} junction without current
concentration, a more general parameter is $I_C/I_B$,
where $I_C$ is the critical current of the structure and $I_B$
is the bulk critical current of the perfect lead.
It seems natural to ask
how is the superconducting flow for intermediate values
of $T_0$ or, more generally, $I_C/I_B$, i.e.,
how is the crossover between the two extreme
limits of superconducting
flow.
This rather fundamental question is of special current relevance, in
view of the recent activity on superconducting mesoscopic structures
(see, for instance, Refs. \cite{c1,c2,furu}).
In the case of a superconducting
point contact, the intermediate regime would correspond
to contact widths not much smaller
than the width of the semiinfinite leads. Alternatively,
in the case of tunneling
barriers, the crossover could be explored by considering
different degrees of transparency at the Fermi level. In the
case of a {\it S-N-S} junction, the intermediate behavior would be
displayed by relatively thin normal metal layers located between
two superconductors.

A preliminary version of some of the results contained
in this article
has been briefly presented in Ref. \cite{lt20}.

\section{Self-consistency and current conservation}

Theoretical studies of weak superconductivity
almost invariably assume that the phase is constant
within the two superconductors.
This is generally a reasonable approximation, since, by
definition, in this regime, $I_C \ll I_B$. As a
consequence, the variation of the phase in the bulk of the superconductor
displayed by current carrying solutions can be safely
neglected in a wide range of length scales.
It is clear that the approximation of
an asymptotically uniform phase cannot be justified
if $I_C$ becomes
comparable to $I_B$,
which will certainly be the case in structures with moderate
or negligibly weakened superconductivity.
The more general situation will be that of a phase
which varies linearly throughout the lead except
in a finite region near the scattering center where it varies
faster.

In order to discuss some general questions related to
self-consistency,
we focus in this section on structures in which the decoupling
between the two sides of a superconductor is due to
one-electron scattering by a barrier or point contact.
The conventional way of generalizing the BCS theory to the
presence of an arbitrary one-electron potential is based on the
Bogoliubov -- de Gennes (BdG) equations \cite{dG}:
\beqa
\left[ \begin{array}{cc}
H_0 & \Delta \\
\Delta^* & -H_0^* \end{array} \right] = \epsilon_n
\left[ \begin{array}{c} u_n \\ v_n \end{array} \right],
\eeqa
where $H_0$ is the one-electron Hamiltonian, $\Delta$
is the gap function, and $[u_n({\bf r}),v_n({\bf r)}]$ and
$\epsilon_n>0$ are, respectively,
the normalized wave function components and the energy of the
quasiparticle $n$. The self-consistency condition for
the gap function is \cite{dG}
\beq
\Delta = V \sum_{n} u_n v_n^* (1 - 2 f_n),
\eeq
where $V$ is the electron-phonon coupling constant and
$f_n=[\exp(\epsilon_n/kT)+1]^{-1}$.
The BdG Hamiltonian can alternatively be written
\beq
H = -\sum_{n\sigma}\epsilon_n \int \! d{\bf r} \, |v_n({\bf r})|^2
+\sum_{n\sigma}
\epsilon_n \gamma_{n\sigma}^{\dagger} \gamma_{n\sigma},
\eeq
where $\gamma_{n\sigma}^{\dagger}$ creates quasiparticle $n$
with spin $\sigma$.
In terms of the quasiparticle operators, the charge and current
density operators are written
\beqa
\rho=&&e\{\sum_{n\sigma}|v_n|^2+\sum_{nm\sigma}(u_n^*u_m-
v_n^*v_m)\gamma_{n\sigma}^{\dagger}\gamma_{m\sigma} \nonumber \\
&& + \sum_{nm\sigma}u_nv_m\sigma\gamma_{m\sigma}\gamma_{n,-\sigma}
+ \sum_{nm\sigma}u_n^*v_m^*\sigma\gamma_{n,-\sigma}^{\dagger}
\gamma_{m\sigma}^{\dagger}\} \\
{\bf j}=&&\frac{e\hbar}{2mi}\{-\sum_{n\sigma}v_n^*\mbox{{\bf D}}v_n
+\sum_{nm\sigma}
(u_n^*\mbox{{\bf D}}u_m+v_n^*\mbox{{\bf D}}v_m)\gamma_{n\sigma}^{\dagger}
\gamma_{m\sigma} \nonumber \\
&& +\sum_{nm\sigma}v_m\mbox{{\bf D}}u_n \sigma \gamma_{m\sigma}
\gamma_{n,-\sigma} - \sum_{nm\sigma} v_m^*\mbox{{\bf D}}u_n^* \sigma
\gamma_{n,-\sigma}^{\dagger}\gamma_{m\sigma}^{\dagger} \},
\eeqa
where $e=-|e|$,
$\sigma=\pm 1$,
and {\bf D} is defined as
$f\mbox{{\bf D}}g \equiv f (\nabla g)\!-\!(\nabla f)g$.
In Eqs. (5) and (6), the
contributions from the condensate and the quasiparticles have been clearly
separated. The quasiparticle contribution can in turn be divided into
a part which conserves the quasiparticle number and a part which
does not. The non-conserving components will not contribute
to the expectation values $<{\bf j}>$ and $<\rho>$ but will play an important
role in the quantum fluctuations of the electronic charge and
current densities.

If one attempts to solve the BdG equations (2) in a given
structure subject to
the boundary condition that the phase takes certain
constant values on each semiinfinite lead,
one generally finds from (6) a nonzero value of the total
current. This general feature can be illustrated by solving exactly a
specific and very important example, namely, that
of a strictly one-dimensional
superconductor (i.e., with only one propagating channel for the
Fermi electrons)
with a barrier of arbitrary transmission $T_0$
at the Fermi level. In this model, the phase is assumed to be
uniform on each side of the barrier. A non self-consistent resolution
of the BdG equations at zero temperature yields the current
\cite{c2,tesis,HKR}
\begin{equation}
I (\Delta \varphi) = \frac{e|\Delta|}{2\hbar} \frac{T_0 \sin(\Delta \varphi)}
{[1-T_0 \sin^2 (\Delta \varphi /2)]^{1/2}}
\end{equation}
where $\Delta \varphi$ is the difference between the phases on each side
of the barrier.
Fig. 1 shows the
current $I(\Delta \varphi)$
for several values of $T_0$.
As the strength of the barrier decreases, the current departs from the
ideal Josephson behavior and its maximum is displaced towards $\pi$.
In particular,
when $T_0$  equals  unity, the current is given by the formula
\cite{c1,tesis}
\begin{equation}
I (\Delta \varphi) = (e|\Delta|/\hbar) \sin(\Delta \varphi/2),
\end{equation}
with $-\pi<\Delta\varphi \le \pi$ and periodicity $2\pi$.
This result is clearly not self-consistent,
since a uniform phase should
be associated with a
vanishing equilibrium current,
at least in the asymptotic region. Actually, a more detailed calculation
reveals \cite{tesis} that the current (6) is localized exponentially
around the barrier
in a region of width
$\pi \xi_0/T_0 \sin(\Delta\varphi/2)$, where $\xi_0=\hbar v_F/\pi|\Delta|$
is the zero-temperature coherence length.
This peculiar feature can be traced back to the existence
of a localized, current-carrying quasiparticle at the interface
\cite{c1,tesis}.
Thus, one finds that the equilibrium
current is nonzero near the scattering center and zero in the
asymptotic region.
In the steady state, this situation clearly involves
a violation of charge conservation.
Below we show that the relation between self-consistency
and current conservation is in fact a general property of the BdG equations.

The time derivative of the charge density operator
can be computed
by applying (4) and (5) to the relation $\dot{\rho}=(1/i\hbar)[\rho,H]$.
The result is
\beqa
\dot{\rho}&=&\frac{e}{i\hbar}\sum_{nm\sigma}\{
(\epsilon_n-\epsilon_m)(v_n^*v_m-u_n^*u_m)\gamma_{n\sigma}^{\dagger}
\gamma_{m\sigma} \nonumber \\
&+&(\epsilon_n+\epsilon_m)\sigma[u_nv_m\gamma_{m\sigma}\gamma_{n,-\sigma}
+u^*_nv^*_m\gamma^{\dagger}_{n,-\sigma}\gamma^{\dagger}_{m\sigma}]\}
\eeqa
which obviously yields $<\dot{\rho}>=0$, as expected from a stationary
scattering description (we have used the properties
$<\gamma^{\dagger}_{n\sigma}\gamma_{n'\sigma'}>=f_n \delta_{nn'}
\delta_{\sigma\sigma'}$ and
$<\gamma_{n\sigma}\gamma_{n'\sigma'}>=0$).
Combining (6) and (9) we obtain for the continuity
equation \cite{lt20,tesis}

\beq
<{\bf \nabla \cdot j}> +<\frac{\partial\rho}{\partial t}>=
\frac{2e}{h}\sum_{n\sigma}
\mbox{Im}\{\Delta^*u_nv_n^*(1-2f_n)\}.
\eeq
By comparing this result with Eq. (3), it becomes
clear that charge conservation is only
guaranteed when the self-consistency condition is satisfied.
In the language of Ref. \cite{kadanoff}, the BCS-BdG theory is
a conserving approximation only for solutions that satisfy the
mean-field equations. It is interesting to note that the
the condensate and
quasiparticle contributions to the electric charge
are not conserved separately, but only
the sum of the two,
and if the description is fully self-consistent. The relation
between self-consistency and current conservation has also
been noticed by Furusaki and Tsukada \cite{furu2}, who have
derived an equation similar to (10) in which condensate and
quasiparticle contributions are however not clearly separated.
This seems to lead to a misinterpretation. Unlike suggested
in Ref. \cite{furu2}, preservation of current conservation is
not achieved in general by merely converting quasiparticle
current into condensate current \cite{furu2,btk}, but by
truly implementing global self-consistency. A good proof of this
assertion is that, within a non self-consistent scheme, the
source term in Eq. (10) is generally nonzero even at zero
temperature,
when no quasiparticles exist.

Before we proceed further a few additional remarks go in place. In one
dimension, Eq. (7) is
incorrect when $T_0$ is not much smaller than unity. In particular, Eq.
(8) is clearly wrong, since no bound quasiparticle should exist in
the absence of a barrier. Of course, the main inconsistency
lies in the very assumption of an existing phase difference, which
cannot be maintained without a scattering obstacle (an abrupt change
in the phase cannot survive the implementation of self-consistency).
It will be seen in the following
section that the appropiate generalization of the concept of
phase difference to structures with arbitrary transparency is the
phase offset, in terms of which
the transparent limit will be quite different from (8).
In studies of superconducting quantum point contacts,
equations which generalize (7) \cite{c2} and (8) \cite{c1,furu} to the presence
of many transverse modes can be found. In these cases, the lack of formal
self-consistency is justified. The localized nonzero current
corresponds to the current in the vicinity of the point contact and the
vanishing of the asymptotic current describes
the widening of the contact into
the reservoir. Therefore, Eqs. (7) and (8), as well as their multimode
generalizations \cite{c1,c2,furu}, are correct as long as
$I_C \ll I_B$. This is the case when the number of propagating modes
in the contact is much smaller than the number of modes in the wide leads.

\section{Study of the crossover}

 From the discussion in the previous section, it is clear
that, in order to achieve a unified view
of the crossover from weak
to strong superconductivity, one must deal with self-consistent,
current conserving
solutions of the BdG
equations in which a nonzero current is
associated with a linearly varying asymptotic phase,
and allow for arbitrary critical currents $I_C \le I_B$.
Unfortunately, the self-consistent resolution of the BdG equations
for arbitrary currents is in general a demanding numerical task.
By contrast, the formalism of Ginzburg-Landau (GL)
provides a relatively simple method to learn
about the global properties of those self-consistent solutions.
Therefore, our goal in this section is to
study the
solutions of the GL equation for a one-dimensional superconductor
in the presence of a delta potential of arbitrary strength.
Specifically, we wish to analyse
the stationary solutions of the free-energy functional
\beq
F=\int dx[|\nabla\psi|^2/\kappa^2-(1-V_0\delta(x))|\psi|^2
+|\psi|^4/2]
\eeq
where
$\kappa=\lambda/\xi$ ($\lambda(T)$ is the penetration depth and $\xi(T)$ is the
temperature-dependent coherence length) and
Abrikosov units are used. In these units, $\lambda(T)$ is the unit of length,
the order parameter $\psi$ is measured in units of $\psi_{\infty}$
(absolute value of the bulk order parameter at zero current), and $(\hbar/m)
(\psi_{\infty}^2/\xi(T)$ is the unit for current.
The complete
crossover between weak and strong superconductivity
will be explored by considering {\it all values
of the scattering strength $g\equiv\kappa V_0$}
ranging from $g$ very large (ideal Josephson
behavior) to $g=0$ (uniform superconductor).
In Eq. (11), $F$ must be understood as the freen energy per unit
area. This model should give a fairly adequate
picture of a quasi-one-dimensional superconductor
(of width $w \ll \lambda,\xi$)
in which a (narrower) point contact \cite{c2}
or a normal metal island has been inserted.
A clean point contact at low temperature
could not be described by (11), since, in the weak superconductivity
limit, this structure yields a current-phase relation of the type (8)
\cite{likharev,c1,kulik} instead of the usual $\sin(\Delta\varphi)$
behavior.
On the other hand, the model (11) is not appropiate for a quantitative
description of tunneling barriers because, in the limit of large
$g$, the repulsive potential $V_0\delta(x)$ yields hard-wall boundary
conditions, which do not correspond to a
GL description of the metal-insulator interface \cite{dG}.
A similar model, with the $\delta$ function replaced by a square barrier,
was studied by Jacobson \cite{jacobson}, who however focussed on the low
current limit. Volkov \cite{volkov} also used a delta function to describe
a {\it S-N-S} junction but only analysed the small current case.

If we
factorize $\psi(x)=R(x)e^{i\varphi(x)}$, the GL equations take the form
\beqa
&&\kappa^{-2} d^2R/dx^2
+[1-V_0\delta(x)](1-j^2/R^4)R
-R^3=0\nonumber\\
&&d\varphi/dx=\kappa j/R^2
\eeqa
where the current density $j$ is a conserved number ($I=jA$).
We are interested in solutions which satisfy the
boundary conditions
\beqa
&&dR(x)/dx = 0 \nonumber\\
&&\varphi(x) = qx \pm \Delta\varphi/2, \;\;\;\mbox{for}\; x
\rightarrow \pm \infty
\eeqa

Current
conservation requires the
product $R^2\varphi'=\kappa j$ to be constant, which can only be
achieved with a {\it nonzero} $q\equiv\kappa j/R^2_\infty$ in the
asymptotic solution. The general solutions for $R$ and $\varphi$
are of the form
\beqa
R^2(x)& = &a+b\tanh^2[\kappa u (x_0+|x|)]\nonumber \\
\varphi(x)& = &qx + \mbox{sgn}(x)\{\arctan[\beta\tanh(\kappa u(x_0
+|x|))] \nonumber\\&& - \arctan[\beta\tanh(\kappa u x_0)]\}.
\eeqa
In Eq. (9), $a(2-a)^2=8j^2$, with $0\leq a \leq 2/3$,
$b=1-3a/2$,
$u=\sqrt{b/2}$, $\beta=\sqrt{b/a}$, and
$x_0$ is obtained from the matching condition at the site of the
delta potential, which gives rise
to the cubic equation
\beq
\sqrt{2b}\beta y_0 (1-y_0^2)
-g(1+\beta^2 y_0^2) = 0,
\eeq
where $y_0\equiv\tanh(\kappa u x_0)$
and thus only the solutions satisfying
$0\leq y_0\leq 1$ are of interest. The solutions turn out to be uniquely
parametrized by the phase offset $\Delta \varphi$, whose general
expression is
\beq
\Delta\varphi=2[\arctan(\beta)-\arctan(\beta y_0)].
\eeq
The resulting curve $j(\Delta\varphi)$
is displayed in Fig. 2.
The inset shows the critical current as a function of $g$.
The Josephson limit is well achieved for $g>8$ while $j_C$
saturates to $j_B=2/3\sqrt{3}=0.385$ as
$g\rightarrow 0$.
For large $g$, one finds the ideal Josephson
behavior, $j=j_C\sin(\Delta\varphi)$, with $j_C=1/2g$ taking small
values. For $g=0$, two types of solutions are obtained. One of them
is entirely expected: for $\Delta\varphi=0$, all currents are possible
ranging from $j=0$ to $j=j_B$. These are
the solutions of the uniform superconductor in which $\varphi'$ and
$R$ take constant values. The second type of solutions
are {\it the solitons of the $\psi^4$
theory defined by} (11)
{\it for arbitrary values of the current $j$} \cite{lt20}.
These kinks
separate two domains in which the phase varies linearly,
\beq
\varphi(x)=qx + \arctan[\beta\tanh(\kappa u x)]
\eeq
with a total phase offset of $\Delta\varphi=2\arctan(\beta)$.
It is interesting to note that, unlike in the $j=0$ case, the phase
offset (which here plays the role of the soliton charge) can
be different from $\pi$.
These solitonic solutions
are equivalent to the saddle-point configurations
which were considered by Langer and Ambegaokar \cite{la}
in their study of the resistive behavior of one-dimensional
superconductors.

Fig. 3 shows the behavior of $R(x)$ and $\varphi(x)$ for $g = 0.2$
and two different values of the current. In Fig. 3a, it is clearly
seen that, for
$j=0.01$, the solitonic solution almost vanishes at $x=0$. For the same
solution,
Fig. 3b shows that the spatial variation of $\varphi(x)$ is almost negligible
except for a step-like feature at $x=0$ (the phase can be shown to vary
in a length scale $j/\kappa$ if $j$ is small).
For $j = 0.35$ (close to $j_B$),
the phase displays a linear increase with $x$ with an offset due to a
faster variation in the vicinity of $x=0$.

An interesting feature of the $j(\Delta\varphi)$
curves which can be clearly observed in Fig. 2 is that,
as the scattering is turned off,
the maximum current is displaced towards lower values of $\Delta\varphi$.
This is in sharp contrast with the behavior shown in Fig. 1 for
the non self-consistent solutions. It has already been noticed that a
superconducting point
contact displays the same behavior as
its propagating channels evolve from low to high transmission
\cite{c2}.
There is of course no contradiction between our results and those obtained
for point contacts,
since the latter apply only in the limit $I_C \ll I_B$, while the low $g$
curves
in Fig. 2 are only relevant in the $I_C \sim I_B$ case.

Baratoff {\it et al.} \cite{bbs} considered a {\it S-}$S'${\it -S} structure in
which $S$ and $S'$ are two dirty superconductors of differing properties.
As a function of the similarity between $S$ and $S'$, they obtained
results which qualitatively resemble those obtained by us. However,
their focus was not in the crossover from weak to strong superconductivity,
but rather in the qualitative modeling of weak links. In particular, they
did not consider the $S=S'$ case and, although the relation
to Ref. \cite{la} is noticed, no association
is made between the branch to the left of the
maximum in the $j\!-\!\Delta\varphi$ curve and
the trivial solutions of the uniform case.
More recently, Kupriyanov \cite{kupri} has studied the
properties of an {\it S-I-S} structure
by means of the Usadel equations, which apply in the dirty limit.
He considers several
values of the barrier transparency and obtains results
which, after a nontrivial
scale transformation (the phase change across the barrier
instead of the phase offset is used as a parameter),
can be shown to be qualitatively similar to
those displayed in Fig. (2). However, in the transparent limit,
no mention is made in Ref. \cite{kupri} of the
relation to the solitonic solutions of Ref. \cite{la}
nor to the uniform solutions, as discussed here by us.

\section{Crossover in long bridges.  Breakdown of the ac Josephson effect.}

So far we have focussed on the relation between the current $j$ and the
phase offset $\Delta\varphi$, which uniquely parametrizes the solutions
of the GL equations (12). However, it is
also convenient to plot the current as a function of the total
phase difference $\chi$ between two reference points.
These two points can be, for instance, the
extremes of a superconductor of length $L$ with an effective barrier
in its center.
A typical case would be that of a narrow bridge connecting two
wide reservoirs through smooth contacts beyond which the phase gradient
can be safely neglected \cite{likharev}. For a given length
$L$, one can compute $\chi$ from the relation $\chi=\varphi(L/2)-
\varphi(-L/2)$. If
$L \gg (\kappa u)^{-1}$,
we can  approximate
\beq
\chi\simeq qL+\Delta\varphi.
\eeq
Since $u\rightarrow 0$ as $j \rightarrow j_C$,
there is a threshold current $j_{th}(L)$ above which
Eq. (18) does not apply.
For $\kappa L \gg 1$, $j_{th}\simeq
j_B(1-27/16\kappa^2L^2)$.
In Fig. 4, the resulting curve $j(\chi)$ is plotted for
$\kappa L=10$.
In such a case $(j_B-j_{th})/j_B \simeq 0.016$.
It can be observed that, for large $g$, the ideal Josephson
behavior is displayed, while, for sufficiently small $g$,
the current becomes a multivalued function of the total phase
$\chi$. The pattern shown in Fig. 4 is actually repeated
periodically with a period of $2\pi$. In the case of $g$ small
it becomes clear from the comparison with Fig. 2 that the
upper branch corresponds to solutions with a linearly
varying
phase ($\Delta\varphi\simeq0$), while the lower branch is
given by the solitonic solutions with a nonzero phase offset.
This feature has also been noticed recently by
Martin-Rodero {\it et al.} \cite{amr}, who have computed
numerically the self-consistent solutions of the BdG equations for
a linear chain coupled to two Bethe lattices at zero temperature.
The discontinuity in the derivative at the top of the $g=0$
curve in Fig. 4 reflects the discontinuous transition from
the uniform to the solitonic branch shown in Fig. 2.
However, this cusp cannot be observed in bridges of finite length since
it always lies above the threshold of validity of Eq. (18).

In Fig. 5 we display
the phase of
the order parameter as a function of the position and the
phase offset for $g=10$ (Josephson limit).
When $\Delta\varphi=\pm\pi$,
the current is zero. This requires an
abrupt jump of $\pm\pi$ at
$x=0$, which is possible because $R(0)=0$ in these
solutions, as can be proven quite
generally.
These are
the phase-slip configurations which permit the existence of the
ac Josephson effect.
As an external driving voltage is applied
between two points on different sides of the junction,
the phase  is forced to vary at a constant rate
and the whole sytem responds
adiabatically by evolving along the continuous set of stationary
solutions.
The existence of these step function solutions
makes it topologically possible for the phase at every
point to increase both monotonically and continuously
with time.
Since $R(0)=0$, the two superconductors are completely
decoupled and $\Delta\varphi=\pi$ is equivalent to
$\Delta\varphi=-\pi$.
As the system is driven by the external
bias through the different values of $\Delta\varphi$ and
reaches the value $\Delta\varphi=\pi$, it automatically
reenters through the topologically equivalent configuration
with $\Delta\varphi=-\pi$ and the phase at the boundary
can continue to increase monotonically. Thus the existence
of the ac Josephson effect relies on the ability of
the system to undergo adiabatic phase-slips under the action
of an external bias. It is interesting to note that, at the
particular value $\Delta\varphi = \pm \pi$, the configuration
of the order parameter is independent of
$g$, since then $R(0)=0$. In particular, it is identical to the
phase-slip configuration in the absence of a barrier, as studied
by other authors (see, for example, Refs. \cite{rieger,kramer}).

At zero current, all points on one
side of the
barrier have the same phase. In particular,
$\varphi(x>0)=\pm\pi/2$ for $\Delta\varphi=\pm\pi$.
By contrast, the solutions with nonzero current have an asymptotic
phase which grows linearly with position, as shown in Eq. (17).
Thus, for sufficiently large $x$, it is not possible to have
$\varphi(x)$ increasing monotonically as $\Delta\varphi$ varies between
the two equivalent configurations with $\Delta\varphi=\pm\pi$. As
a consequence, the system cannot respond
adiabatically
to a constant voltage being applied at points that are
sufficiently distant from the junction. The only choice for
the system will be to undergo nonadiabatic, fluctuating processes
of the type studied by Langer and Ambegaokar \cite{la} (albeit
with $g\neq0$), which will originate a resistive behavior.
The threshold for this type of response is given by the condition
\beq
\left.\frac{\partial\varphi(x_b)}{\partial\Delta\varphi}
\right
|_{\Delta\varphi=\pi}=0
\; \; \; \mbox{at}  \; \; \Delta\varphi=\pi
\eeq
If one identifies $x_b=L/2$ and $\varphi(x_b)=\chi/2$, this is also
the condition for
the onset of bivaluedness
in the $j(\chi)$ curve of Fig. 4, which requires
$\partial\chi/\partial j=0$ at $\chi=\pi$ (note that,
$\partial j/\partial \Delta\varphi\neq 0$ at
$\Delta\varphi=\pi$). Thus, if the electrodes are applied at
points $|x|>x_b$,
there is a breakdown of the ac Josephson effect
due to the fundamental inability of the system to respond
adiabatically to that particular type of external constraint.

Let us estimate the breakdown length $x_b$.
For $g$ large, one can show that $x_b\simeq g/\kappa$.
We notice at this point that the value of the parameter
$g$ can be adjusted to a realistic setup by exploiting the
relation
\beq
g=1.30\cdot(j_B/j_C),
\eeq
which applies
in the Josephson regime, and noting that $j_B/j_C=I_B/I_C$.
We have considered explicitly four types of structures which are
known to display a standard $\sin(\Delta\varphi)$ behavior
in the
Josephson limit
for $T$ close to $T_c$: (a) a tunnel junction with average
transmission $T_0$ for the Fermi electrons, (b) a clean point
contact with average transmission $T_0$, (c) a narrow bridge
between two superconductors
made of a dirty normal metal of length $L$ and coherence
length $\xi_N$ at $T\simeq T_c$, and (d) a {\it S-N-S} structure
without current concentration ($N$ and $S$ have the same width).
Cases (a) and (b) fall within the same category in the GL
limit, with an expression $I_C=\pi \Delta^2(T)/4eR_Nk_BT$ for the
critical current \cite{likharev}.
Noting that, for $T$ close to $T_c$, the gap
function and the order parameter are related by \cite{abrikosov}
$\psi=0.326\,\sqrt{n}\Delta/k_BT_c$, where $n$ is the electron
number density, we arrive at
\beq
g^{-1}\simeq 2.0 \; T_0 (\xi(T)/\xi_0).
\eeq
For case (c), the critical current is \cite{likharev}
$I_C=(4\Delta^2(T)/\pi eR_Nk_BT)(L/\xi_N)\exp(-L/\xi_N)$, if
$L \gg \xi_N$. Thus one obtains
\beq
g^{-1}\simeq 3.23 \;T_0 (\xi(T)/\xi_0)(L/\xi_N)\exp(-L/\xi_N).
\eeq
For a {\it S-N-S} structure without current concentration, the critical
current is \cite{abrikosov1} $I_C=A(e\hbar n/2m)(|T-T_c|/T_c)
(\xi_N/\xi^2(T))\exp(-L/\xi_N)$. As a consequence,
\beq
g^{-1}\simeq 10.6 \;(\xi_N/\xi(T))\exp(-L/\xi_N).
\eeq
Shifting to real units, we arrive at the relations
\beqa
x_b & \simeq & 0.50 \; \xi_0/T_0
\mbox{\hspace{3cm}(a),(b)} \nonumber \\
x_b & \simeq & 0.31 \; (\xi_0\xi_N/T_0L)e^{L/\xi_N}
\mbox{\hspace{1cm}(c)}
\nonumber \\
x_b & \simeq & 0.094 \; (\xi^2(T)/\xi_N)e^{L/\xi_N}
\mbox{\hspace{0.8cm}(d)}
\eeqa
for the maximum distance at which a constant voltage can be applied
in order to observe the ac Josephson effect.

\section{Conclusions}

We have studied the nature of
the crossover from ideal Josephson behavior between two weakly
coupled superconductors to bulk
superconducting flow in a perfect superconducting lead.
We have argued that a
self-consistent resolution of the BdG equations
is mandatory in a microscopic study of the crossover and have
proved that charge conservation is only guaranteed when
the requirement of self-consistency is satisfied.
We have performed a study of the crossover by
solving exactly
the Ginzburg-Landau equation for a
one-dimensional superconductor in the presence of a delta potential
of arbitrary strength. The pairs of Josephson solutions with equal
current have their scattering free counterparts in the pairs
formed by a uniform
and a solitonic solution. This relation has allowed us to understand
some aspects
of the  multivalued current-phase relation in
narrow bridges.
The complete knowledge of the set of stationary solutions for
different values of the scattering strength $g$ has helped us to gain
a more detailed understanding of the
adiabatic response to a constant external
bias, which has been shown to rely
on the feasibility of
adiabatic phase-slips.
If a voltage is applied at points which are sufficiently
far from the junction, there is a breakdown of the Josephson
effect due to the intrinsic impossibility of changing adiabatically
the phase
at a distant point in a continuous and monotonic manner.

\acknowledgments
The authors would like to thank P. Ao, C.W.J. Beenakker,
F. Flores, P. Goldbart, F.
Guinea, A. J. Leggett, A. Levi Yeyati, and A. Martin-Rodero for
useful discussions at various stages of this work. This project
has been supported by CICyT, Project no. MAT91-0905.

\begin{figure}
\caption{Current $j$  as function of  the
phase offset $\Delta\varphi$  for a non-self-consistent solution
The curves
labeled {\it a, b, c}, and $d$ are for the cases
$T_0 = 0.999, 0.99, 0.9$ and $0.4$, respectively.}
\end{figure}
\begin{figure}
\caption{Current $j$  as function of the
phase offset $\Delta\varphi$.
The curves are
labeled {\it a, b, c} and $d$ for the cases
$g = 0, 0.5, 3$ and $10$, respectively. Inset: critical
current $j_C$ versus scattering strength $g$; solid line gives the
the exact result and dotted line corresponds to
the Josephson limit $1/2g$.}
\end{figure}
\begin{figure}
\caption{The amplitude (a) and the phase (b) of the order parameter plotted
as a function of position in the $g=0.2$ case ($\kappa=1$),
for values of the current
$j=0.01$ (curves labeled $a$) and $j=0.35$ (curves labeled $b$).}
\end{figure}
\begin{figure}
\caption{
Same as Fig. 2, for the total phase difference $\chi$ between the
extremes of a superconductor of length $L=10$.}
\end{figure}
\begin{figure} \caption{The phase of the order parameter is plotted as a
function of the position $x$ ($\kappa=1$) and the phase offset $\Delta\varphi$
for the case $g=10$.}
\end{figure}
\end{document}